\documentclass[journal]{IEEEtai}

\usepackage[colorlinks,urlcolor=blue,linkcolor=blue,citecolor=blue]{hyperref}

\usepackage{color,array}

\definecolor{reviewcolor}{rgb}{0,0,0}
\newcommand{\newtext}[1]{{\color{reviewcolor}#1}}

\usepackage{graphicx}

\usepackage{amssymb}
\usepackage{algorithmic}
\usepackage{multirow}
\usepackage[cmex10]{amsmath}
\usepackage{booktabs}

\begin{document}

\title{Teacher-Student Structure for Domain Adaptation in Ensemble Audio-Visual Video Deepfake Detection} 

\author{Elham Abolhasani,
Maryam~Ramezani*,
			and~Hamid~R.~Rabiee*
			\IEEEcompsocitemizethanks{
            *: corresponding authors\protect\\
            E-mail:\{elham.abolhasani199,maryam.ramezani,rabiee\}@sharif.edu\protect\\
            Department
				of Computer Engineering, Sharif University of Technology.
			}
			% \thanks{Manuscript received July 27, 2025.}
            }

% \markboth{Journal of IEEE Transactions on Artificial Intelligence, Vol. 00, No. 0, Month 2025}
% {E. Abolhasani \MakeLowercase{\textit{et al.}}: Teacher-Student Structure for Domain Adaptation in Ensemble Audio-Visual Video Deepfake Detection}

\maketitle

\begin{abstract}
The rapid advancement of generative AI models is leading to more realistic deepfake media, encompassing the manipulation of audio, video, or both. This raises severe privacy and societal concerns. Numerous studies in this area have yielded promising intra-domain results; however, these models frequently exhibit decreased efficacy when faced with data from dissimilar domains. Consequently, recent deepfake detection approaches focus on enhancing the generalization ability through multiple techniques that incorporate all input modalities, including audio, images, and their interactions. In this regard, we propose the EAV-DFD method, a generalized deep ensemble audio-visual model (EAV-DFD) combined with a domain adaptation mechanism utilizing a teacher-student framework to enhance the model’s ability to perform and generalize effectively across unseen domains. To evaluate the model's performance, we used the FakeAVCeleb dataset as the primary domain and the DFDC, Deepfake\_TIMIT, and \newtext{PolyGlotFake} datasets as an unseen domain. Our experimental results demonstrate that the proposed framework is efficient in domain adaptation,  improving AUC performance of the model by 4.09\%, 17.94\%, and \newtext{0.5\%} on \newtext{three} unseen datasets, using only a small portion of them to train the student model. This leads to a novel deepfake detection model capable of adapting to new domains and interpreting which modality has been manipulated, highlighting the potential of our approach for real-world applications.
\end{abstract}

\begin{IEEEImpStatement}
Advances in AI and the abundance of data have driven the rapid evolution of deepfake generation methods, enabling the creation of high-quality content. Due to growing concerns over the potential misuse of these new-generation methods, there is a critical need for advanced and robust detection models. This work introduces a novel deepfake detection approach that integrates audio-visual ensemble learning with a teacher-student framework, enabling continuous adaptation to evolving generation methods. The proposed model can learn new deepfake methods without forgetting prior knowledge, using a small number of videos from both old and new domains. Experimental results demonstrate that our model also performs well even with unimodal input when only audio or visual modality is available.  Furthermore, our model's architecture and training strategy allow for explainable predictions by providing separate probability outputs for each modality. These features make this model well-suited for real-world scenarios due to its adaptability under different conditions.
\end{IEEEImpStatement}


\section{Introduction}

\IEEEPARstart{T}{he} swift progression of deepfake generation models has led to the rise of highly realistic synthetic content with either or both visual and audio modalities manipulation. Although these techniques have opened new frontiers in media production and applications, such as video games, animations, and virtual reality, their malicious use can also pose significant threats to privacy, security, and public trust.
For instance, on the eve of the 2024 US presidential election, a deepfake video shared by Elon Musk on the social media platform X mimics the voice of Vice President Kamala Harris, attributing to her statements she did not make. This video raises critical questions about the ethical implications of AI-generated content and sparked debate about the power of artificial intelligence to mislead, especially in the context of politics. In response to these concerns, it is crucial to develop advanced deepfake detection models that aim to ensure the trustworthiness of digital media. 
Therefore, researchers have developed numerous techniques to detect deepfakes. Many of these methods rely on a unimodal approach, using either audio or visual information to detect forged content
\cite{yang2019exposing, singh2021detection}. 

With the growing quality of deepfake media and the uncertainty around which modality is being manipulated, researchers have also focused on multimodal deepfake detection. This approach leverages more comprehensive information, including audio, visual, and their interrelationship, to enhance the performance and generalization of detectors
\cite{zhou2021joint}.
But in real-world scenarios, there are instances where either the audio or visual modality is missing. However, most existing audio-visual detection models are impractical in these scenarios.
In addition, recently developed unimodal and multimodal deepfake models have achieved good performance on test data from the same domain as their training data; however, their performance often drops significantly on unseen data from different domains. This poses a significant challenge to generalization in this field. To tackle this issue, researchers try to use different methods, like data augmentation, training models only on real datasets
\cite{cozzolino2023audio}
and extracting intrinsic and abstract features
\cite{boccignone2022deepfakes}.
Although these techniques help models generalize better, the realism and variety of new deepfakes necessitate adaptation methods to enhance performance and robustness in unseen domains.
In this work, we present a novel generalized deep ensemble audio-visual deepfake detection model (EAV-DF) with a teacher-student structure for adapting to new domains. To develop a generalized model that performs well in real-world scenarios, we applied several methods to enhance the model's generalization ability. We employed a structure to adapt the model to newly generated content.

\begin{figure*}[]
  \begin{center}
    \includegraphics[width=0.61\textwidth]{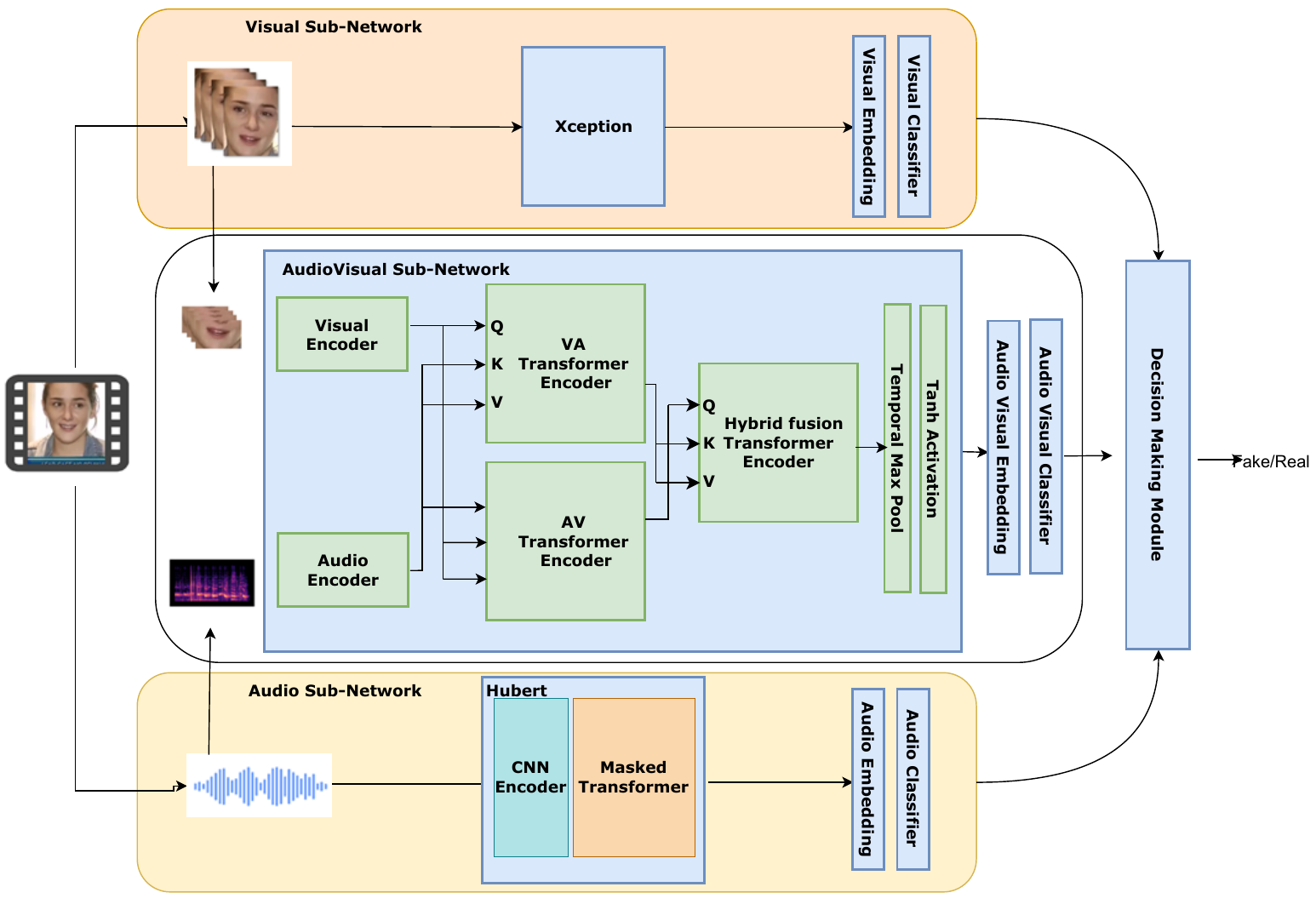}
    \caption{Overview of the proposed EAV-DF model. We first segment the video into smaller clips, where each clip is considered an input to the model. We then prepare the input for the visual sub-network by extracting facial crops from video frames. For the audio-visual sub-network, we use lip region crops along with Mel-spectrogram features derived from the audio. The audio sub-network additionally processes raw audio, where a CNN Encoder extracts features that are subsequently used by the Masked Transformer component. Then, each sub-network generates embeddings, which are classified by three distinct classifiers. Finally, a decision-making module integrates the predictions to determine the final label.}
    \label{fig1}
  \end{center}
\end{figure*}

As shown in Figure
\ref{fig1}, 
The proposed EAV-DFD model is an ensemble of audio, video, and audio-visual sub-networks, along with a decision-making module to comprehensively utilize the input information from multiple perspectives. The unimodal networks process the audio and visual modalities of input separately to make predictions based on the artifacts in each modality, thereby enhancing detection in the decision-making module. Additionally, the presence of these networks can handle scenarios where the input consists of only a single modality. In an audio-visual network, after extracting feature embeddings of each modality by the corresponding encoder, cross-attention transformers integrate these embeddings to exploit inconsistencies and interactions between audio and visual data. Finally, a decision-making module combines the results of the three networks to make the final prediction of the model.
After designing the model architecture, we first train each sub-network separately to ensure a well-initialized starting point. We then fine-tune the model on the FakeAVCeleb dataset and apply several data augmentation techniques to enhance the model's generalization ability. We also employ a teacher-student framework to develop a model that can dynamically adapt to new domains. Therefore, the teacher model, which is trained on the initial domain, transfers its learned knowledge to the student model using specialized cost functions. This enables the student model to adapt to unseen domains while maintaining its performance in the primary domain. This approach allows for the student model to effectively bridge the gap between the source and target domains, ensuring consistent performance across both.

The main contributions of our paper can be summarized as:

\begin{itemize}
  \item	We propose a novel deepfake detection architecture capable of identifying forgery clues and inconsistencies across both audio and visual modalities. We train our model on the primary dataset using a proposed loss function including the binary cross-entropy and contrastive loss functions.
  
  \item Our model employs an ensemble structure, integrating an audio-visual network alongside two distinct unimodal networks. This design enhances detection effectiveness and ensures robust performance even when one modality is missing.
  
  \item In addition to multimodal datasets, we leverage unimodal deepfake and real video datasets to pre-train each sub-network (Details in Table I of the Supplementary Materials). This enables better generalization and speeds up the model’s training on the primary dataset.
  
  \item We introduced a teacher-student framework to facilitate the model's adaptation to unseen data of new domains. This approach involved designing specific loss functions that guide the student model, under the teacher's supervision, to perform well on both main and unseen domains.
  
  \item Comprehensive evaluations are conducted on four datasets that demonstrate the generalization and adaptability of our proposed method. 
\end{itemize}
% The rest of the paper is organized as follows: Section \ref{sec:relatedworks} reviews the related works. Section \ref{sec:proposedmethod} introduces the proposed EAV-DF method and details the teacher-student structure for domain adaptation. Section \ref{sec:dataset} outlines the utilized datasets and implementation details. Section \ref{sec:experiments} reports the experimental results and provides some analyses. The conclusion is made in Section \ref{sec:conclusion}.
\section{Related Works}
\label{sec:relatedworks}

The spread of deepfakes and the ease of creating them have driven research into detection methods, which are categorized into unimodal and multimodal approaches based on the modalities they utilize for detection.  In this section, we also mention works that use methods to learn new, unseen data after training.

\subsection{Unimodal Deepfake Detection}
The early efforts of researchers in this field focused on detection based on hand-crafted features. For visual deepfakes, the Headpose model
\cite{yang2019exposing}
extracted head orientation and position estimated from 2D landmarks and
\cite{caldelli2021optical} 
used CNNs to distinguish motion dissimilarities in videos by exploiting optical flow fields. For audio deepfakes, some works like
\cite{singh2021detection}
used MFCC and Mel-spectrogram features of audio for detection. 
After that, models are gradually moved towards using deep learning methods for more effective feature extraction.  So, they employed different architectures for detection, such as Capsule Networks
\cite{nguyen2019capsule}, Xception model
\cite{rossler2019faceforensics++},
Transformers
\cite{wang2022m2tr} and even RNNs
\cite{guera2018deepfake}.
With the growing realism of synthetic data, models also focused on extracting more intrinsic features for robust detection. Therefore, some methods tried to use biometric features like heart rate, physiological behavior
\cite{boccignone2022deepfakes}
and lip movement
\cite{haliassos2021lips}.
Another example uses a speech emotion recognition network for extracting audio features \cite{conti2022deepfake}.
These approaches have shown promise but can still be improved. Enhanced detection of synthesized videos, mainly those newly manipulated in both audio and visual aspects, can be achieved through multimodal analysis.

\subsection{Multimodal Deepfake Detection}

Multimodal detection is another approach to detecting deepfakes and overcoming the difficulties of unseen forgeries. By leveraging both audio and visual modalities simultaneously, these models can detect artifacts within each modality and inconsistencies between them.
\cite{zhou2021joint}
is the first articles to demonstrate the effectiveness of considering audio-visual synchronization features in deepfake detection, they exploit the relationship between audio and visual modalities to detect fake samples. Following this work,
\cite{chugh2020not} has used contrastive learning to detect deepfakes based on the similarity between audio and visual information.
Some other works have focused on learning audio and visual signals jointly. In this context, different fusion techniques have been explored. For instance, AVFF 
\cite{oorloff2024avff}
introduced a two-stage cross-modal method to learn the audio-visual correspondences by a dual objective of contrastive learning and autoencoding, supplemented by an audio-visual complementary masking and fusion strategy. Additionally, some researchers have employed transformers with cross-attention mechanisms to achieve this goal. AVoiD-DF
\cite{yang2023avoid}
focused on exploring audio-visual inconsistency at both temporal and spatial levels, followed by a joint decoder with a cross-attention mechanism for jointly learning audio-visual inconsistency. Another example is AVT²–DWF
\cite{10609529}
which proposed a framework to address the subtle spatial variances and temporal consistencies within video content. This model highlights the attributes of each modality using face transformer and audio transformer encoders, and employs a dynamically weighted fusion technique to extract common attributes from the audiovisual modalities.
In another line of work,
\cite{yu2023pvass}
has shown that audio-visual correspondences can be learned from real videos. This self-supervised auxiliary method can improve the accuracy of detecting audio-visual inconsistencies in multimodal deepfakes. The authors of
\cite{liu2023magnifying}
also introduced a transformer module to enhance temporal intra-modal artifacts within the audio and video modalities. Then, they designed a distribution difference module to align multimodal information, further magnifying cross-modal inconsistency adaptively. Since most multimodal deepfake detection models are unable to identify unimodal deepfakes, several studies have explored the use of ensemble methods to address this issue
\cite{zhang2023joint, usmani2025spatio, hashmi2025avtenet}.
However, these models tend to improve performance and generalization ability, but none of the above methods can learn new data after training. We propose an ensemble audio-visual deepfake detection model that can adapt to new unseen data after training using a teacher-student framework to address the above issues.

\subsection{Methods for Adapting Models to New Domain Data}
Researchers are developing generalizable models to tackle the issue of deepfake technologies advancing quickly. They employ domain adaptation techniques to learn about new forgery artifacts created by various methods.
Given this context, some studies employed incremental and transfer learning approaches
\cite{lee2021tar, khan2021video}
to facilitate more effective learning from new data.
Some others tried to use the teacher-student framework for domain adaptation.
For example, \cite{kim2021fretal}
introduced a Fretal knowledge distillation-based method to train a student model on the target domain from a pre-trained teacher model on the source domain to transfer representations. The authors of 
\cite{yang2022confidence}
also proposed a CDC (Contrastive Distillation Calibration) framework that incorporates a dual-teacher module, where each teacher is trained on a specific type of forgery. They then employed a dynamic weight Kullback-Leibler (KL) divergence loss and label smoothing approach to distill knowledge from the dual-teacher block to the student model. 
However, these methods tried to adapt models to unseen data and enhance their performance on them, but their basic models are visual-only networks with simple architectures. So they are not powerful enough to learn more complex samples that contain high-quality fake frames or only audio-manipulated ones. Motivated by
\cite{kim2021fretal, yang2022confidence}
to address these issues, we introduce a teacher-student framework that enables our proposed ensemble audio-visual model to adapt and learn from unseen data after model training.

\section{Proposed Method}
\label{sec:proposedmethod}
As summarized in Table \ref{tab:models_comparison}, our model holds a distinct position among existing deepfake detection methods discussed in the related work section. It is an ensemble model of unimodal and multimodal components that utilizes a teacher-student framework for domain adaptation. This section outlines the network architecture and details the training process of the teacher model, along with the teacher-student framework used for domain adaptation.

\begin{table}[h]
\caption{Comparison of Deepfake Detection Approaches in Terms of Modality and Domain Adaptation Support}
\label{tab:models_comparison}
\centering

\resizebox{\columnwidth}{!}{%
\begin{tabular}{c c c c}
\hline
\textbf{Models} & \textbf{Unimodal} & \textbf{Multimodal} & \textbf{Domain Adaptation} \\
\hline
Xception \cite{chollet2017xception}, LipForensics \cite{haliassos2021lips} & \checkmark & & \\
\hline
MDS \cite{chugh2020not} & &  \checkmark & \\
\hline
AVA-CL \cite{zhang2023joint}, AVTENet \cite{hashmi2025avtenet}, \\ AVFakeNet \cite{ilyas2023avfakenet}, AVT²–DWF \cite{10609529}, \\
STKD-VViT \cite{usmani2025spatio}, SS-AVD \cite{zhang2025lightweight}&   \checkmark& \checkmark & \\
\hline
Fretal  \cite{kim2021fretal} , CDC \cite{yang2022confidence} & \checkmark & & \checkmark \\
\hline
This paper & \checkmark & \checkmark  & \checkmark \\
\hline
\end{tabular}%
}
\end{table}

\subsection{Network Architecture}

As shown in Figure \ref{fig1}, our proposed forgery detection model consists of three key components:  \textbf{Visual Network} that extracts remained artifacts in frames,  \textbf{Audio Network} which finds an informative embedding for audio signal, and  \textbf{Audio-Visual Network} that explores inconsistencies between the audio and visual modalities to learn a joint embedding through their interaction. The embeddings extracted by each sub-network are subsequently passed through dedicated MLP classifiers. Finally, a decision-making module aggregates the outputs from the three sub-networks to determine the final prediction. We train our model on the primary domain dataset, establishing this base model as the teacher network for the domain adaptation phase. After that, to develop an architecture capable of learning new data after training, we employ a teacher-student framework to adapt our model and improve its performance on unseen domain datasets. The symbols used in this paper are summarized in Table \ref{tab:notation}. 
\begin{table*}[t]
    \centering
    \caption{Notation Table}
    \label{tab:notation}
    \resizebox{0.75\textwidth}{!}{%
    \begin{tabular}{c c}
        \hline
        \textbf{Notation} & \textbf{Description} \\
        \hline
        $A/V/AV/VA$ & Audio, Visual, Audio-Visual, Visual-Audio \\
        $C$ & Number of frames in each clip \\
        $V^{(c)}$ & A sequence of $C$ video frames \\
        $MFCC$ & Mel-Frequency Cepstral Coefficients \\
        $H/W$ & Height, Width \\
        $L_t$ & Selected layers of transformers for applying KL divergence loss \\
        $P_a, P_v, P_{\text{av}}$ / $L_a, L_v, L_{\text{av}}$ & Output probability / label of each sub-network \\
        $h_a, h_v, h_{\text{av}}$ / $f_a, f_v, f_{\text{av}}$ & Audio, visual, and audio-visual sub-networks without / with their associated classifiers \\
        $h_{\text{T}(m)} / h_{\text{S}(m)}$ & Sub-networks $(m \in \{a, v, av\})$ in teacher / student model without their associated classifiers \\
        $f_{\text{T}(m)} / f_{\text{S}(m)}$ & Sub-networks $(m \in \{a, v, av\})$ in teacher / student model with their associated classifiers \\
        $P/L$ & Final fake probability / label of ensemble model \\
        \hline
    \end{tabular}%
    }
\end{table*}
\subsection{Audio-Visual Sub-Network}
The main part of our model is an audio-visual sub-network, which is designed to exploit interactions between audio and visual data, similar to the approach used in
\cite{kadandale2022vocalist} 
for audio-visual synchronization classification. We provide a sequence of cropped mouth frames and normalized Mel-spectrogram features of audio as inputs to this sub-network. First, these inputs are fed into the corresponding CNN encoders, which consist of 3D/2D convolutional layers with residual skip connections. Then we employ a cross-modal network composed of three transformers with a cross-attention mechanism to track the correlation between audio and visual signals. 
\newtext{
Each transformer is composed of 4 layers, 8 attention heads, and a hidden unit size of 200. The first two transformers take extracted features from encoders as inputs, and for each transformer, the query is derived from one modality. In contrast, the key and value pairs are drawn from the second modality. This means that the VA transformer learns to understand the visual input by focusing on parts of the visual features that correspond most closely to the audio features. In contrast, the AV transformer performs the opposite task. Then, the fusion transformer serves the output of the AV transformer as query and the output of the VA transformer as key and value to make up combined features.}
Finally, we apply a temporal max-pooling layer followed by a tanh activation function to select more prominent features and reduce the dimension. 

\subsection{Audio and Visual Sub-Network}
In addition to the audio-visual network, we employ distinct audio and visual networks in our model to enhance the learning of modality-specific representations. For the visual network, while the audio-visual network focuses on temporal information across frames, we used a simple Xception \cite{chollet2017xception} network that emphasizes the detailed features of individual frames, capturing essential visual information. On the audio side, to learn a robust audio representation, we utilized the HuBERT \cite{hsu2021hubert} model. This model comprises two components: a CNN encoder that processes raw audio and extracts low-level acoustic features, followed by a masked transformer that subsequently operates on the encoded features. The mean of the final hidden states from the transformer provides a global representation used for classification.

\subsection{Decision-Making Module}

According to Algorithm %\ref{alg:decision_making}%
1, we feed one video clip to all three sub-networks to extract the corresponding embeddings ($h_a, h_v, h_\text{av}$), and then we use separate MLP classifiers to find the label of each sub-network. Since video manipulation can occur in any of the modalities, we set the final label of our network as fake if at least one of the sub-networks detects a fake label. The final probability, in this case, is computed as the mean of the probabilities from only the fake-detecting sub-networks. Conversely, if no manipulation is detected, the average across all probabilities is used. Finally, the overall label for a video is determined by processing $N$ clips, averaging their final probabilities, and applying a threshold-based decision rule.
% \vspace{0.25cm}
%  {\small
% \begin{algorithmic}[1]
% \renewcommand{\algorithmicrequire}{\textbf{Input:}}
% \renewcommand{\algorithmicensure}{\textbf{Output:}} 
% \hline
% \text{\textbf{Algorithm 1} Decision Making Algorithm}  \\
% \hline 

% \REQUIRE C frame of Video $V^{(c)}$: $[C, H, W]$, Threshold $T=0.5$
% \ENSURE Label $L$, Prob $P$

% \STATE Initialize $L \gets 0$

% \STATE \text{/* Find Label of each Sub-Networks}
% \STATE \hspace{6pt} \text{Label 1 means Fake Label and }
% \STATE \hspace{6pt} \text{Label 0 means Real Label */}

% \STATE $P_\text{av}, P_a, P_v$ \gets EAVNetwork($V^{(c)}$)

% \STATE \textbf{if} $P_\text{av} > T$ \textbf{then} $L_\text{av} \gets 1$ \textbf{else} $L_\text{av} \gets 0$
% \STATE \textbf{if} $P_a > T$ \textbf{then} $L_a \gets 1$ \textbf{else} $L_a \gets 0$
% \STATE \textbf{if} $P_v > T$ \textbf{then} $L_v \gets 1$ \textbf{else} $L_v \gets 0$

% \STATE {\text{/* Find ensemble label and probability */}}
% \STATE $L \gets OR(L_\text{av}, L_a, L_v)$
% \STATE $\textbf{if}$   $L == 1$
% \STATE \hspace{1em} $P \gets \frac{L_\text{av}*P_\text{av} + L_a*P_a + L_v*P_v}{sum(L_\text{av}, L_v, L_a)}$
% \STATE $\textbf{else} $
% \STATE \hspace{1em} $P \gets mean(P_\text{av}, P_a, P_v)$

% \RETURN $L$, $P$ \\[1mm]
% \hline
% \label{alg:decision_making}
% \end{algorithmic}
% }
\vspace{0.25cm}
{\small
\begin{minipage}{\linewidth}
\hrule height 0.4pt
\vspace{1mm}

\noindent\hspace*{1.2em}\textbf{Algorithm 1} Decision Making Algorithm

\vspace{1mm}
\hrule height 0.4pt
\vspace{1mm}

\begin{algorithmic}[1]
\renewcommand{\algorithmicrequire}{\textbf{Input:}}
\renewcommand{\algorithmicensure}{\textbf{Output:}}

\REQUIRE C frame of Video $V^{(c)}$: $[C, H, W]$, Threshold $T = 0.5$
\ENSURE Label $L$, Prob $P$

\STATE Initialize $L \gets 0$

\STATE /* Find Label of each Sub-Networks
\STATE \hspace{6pt} Label 1 means Fake Label and
\STATE \hspace{6pt} Label 0 means Real Label */

\STATE $P_{\mathrm{av}}, P_a, P_v \gets \mathrm{EAVNetwork}(V^{(c)})$

\STATE \textbf{if} $P_{\mathrm{av}} > T$ \textbf{then} $L_{\mathrm{av}} \gets 1$ \textbf{else} $L_{\mathrm{av}} \gets 0$
\STATE \textbf{if} $P_a > T$ \textbf{then} $L_a \gets 1$ \textbf{else} $L_a \gets 0$
\STATE \textbf{if} $P_v > T$ \textbf{then} $L_v \gets 1$ \textbf{else} $L_v \gets 0$

\STATE /* Find ensemble label and probability */
\STATE $L \gets OR(L_{\mathrm{av}}, L_a, L_v)$
\STATE \textbf{if} $L == 1$
\STATE \hspace{1em} $P \gets \frac{L_{\mathrm{av}} * P_{\mathrm{av}} + L_a * P_a + L_v * P_v}{\mathrm{sum}(L_{\mathrm{av}}, L_v, L_a)}$
\STATE \textbf{else}
\STATE \hspace{1em} $P \gets \mathrm{mean}(P_{\mathrm{av}}, P_a, P_v)$

\RETURN $L$, $P$
\end{algorithmic}

\vspace{1mm}
\hrule height 0.4pt
\label{alg:decision_making}
\end{minipage}
}

\subsection{Loss Functions For Training Teacher Model}

For the training of the teacher model, we first adopt a separate cross-entropy loss function for the classifier output of each sub-network, defined as:

\begin{equation} \label{eq:1}
{\small
\begin{split}
L_{B C E} = \sum_{m \in\{a, v, a v\}} & -\frac{1}{N}  \sum_{i=1}^N \bigg( y_{m, i} \log \sigma\left(f_m\left(x_{m, i}\right)\right) +  \\
& \left(1-y_{m, i}\right) \log \left(1-\sigma\left(f_m\left(x_{m, i}\right)\right)\right) \bigg)
\end{split}
}
\end{equation}

Here, \( N \) denotes the total number of training samples, $y_{m, i}$ shows the label of the $i$th data sample, and $m \in \{a, v, av\}$ refers to the audio, visual, and audio-visual sub-networks $f_m$.

To ensure a more effective distinction between real and fake samples in the audio-visual sub-network, we also incorporate a contrastive loss with margin M for the output of the AV and VA transformers based on Equation \ref{eq:2} to make them close for the real samples and far for the fake ones.

\begin{equation} \label{eq:2}
{\small
\begin{gathered}
L_C=\frac{1}{N} \sum_{i=1}^N\left(y_{a v, i}\right)\left(d_i\right)^2+\left(1-y_{a v, i}\right) \max \left({M}-d_i, 0\right)^2 \\ d_i=\left\|h_v-h_a\right\|_2
\end{gathered}
}
\end{equation}

Finally, the overall training loss is defined as $L_{Teacher} = L_{B C E} + \alpha L_C$ where the weighting factor $\alpha=0.005$ was determined through validation.

% Then, we consider the total loss function for the training as the sum of the cross-entropy and the contrastive loss, which is controlled by the hyperparameter $\alpha$, as shown in Equation \ref{eq:3}.
% \newtext{
%  The optimal value $\alpha=0.005$ was determined through validation.
% }

% \begin{equation} \label{eq:3}
% \begin{gathered}
% L_{Teacher} = L_{B C E} + \alpha L_C
% \end{gathered}
% \end{equation}

\subsection{Domain Adaptation using Teacher-Student Framework}
In this study, we aim to obtain a generalized model that is robust to changes and new deepfakes. But as the variety and quality of deepfake generation methods increase, using generalization techniques is not enough, and the model needs to be adapted to perform well on new data with distribution shifts. To achieve this, we propose a domain adaptation mechanism based on a teacher-student framework to build a deepfake detector model that is efficient on unseen samples that differ from the training data.

\begin{figure*}[]
  \begin{center}
    \includegraphics[width=0.9\textwidth]{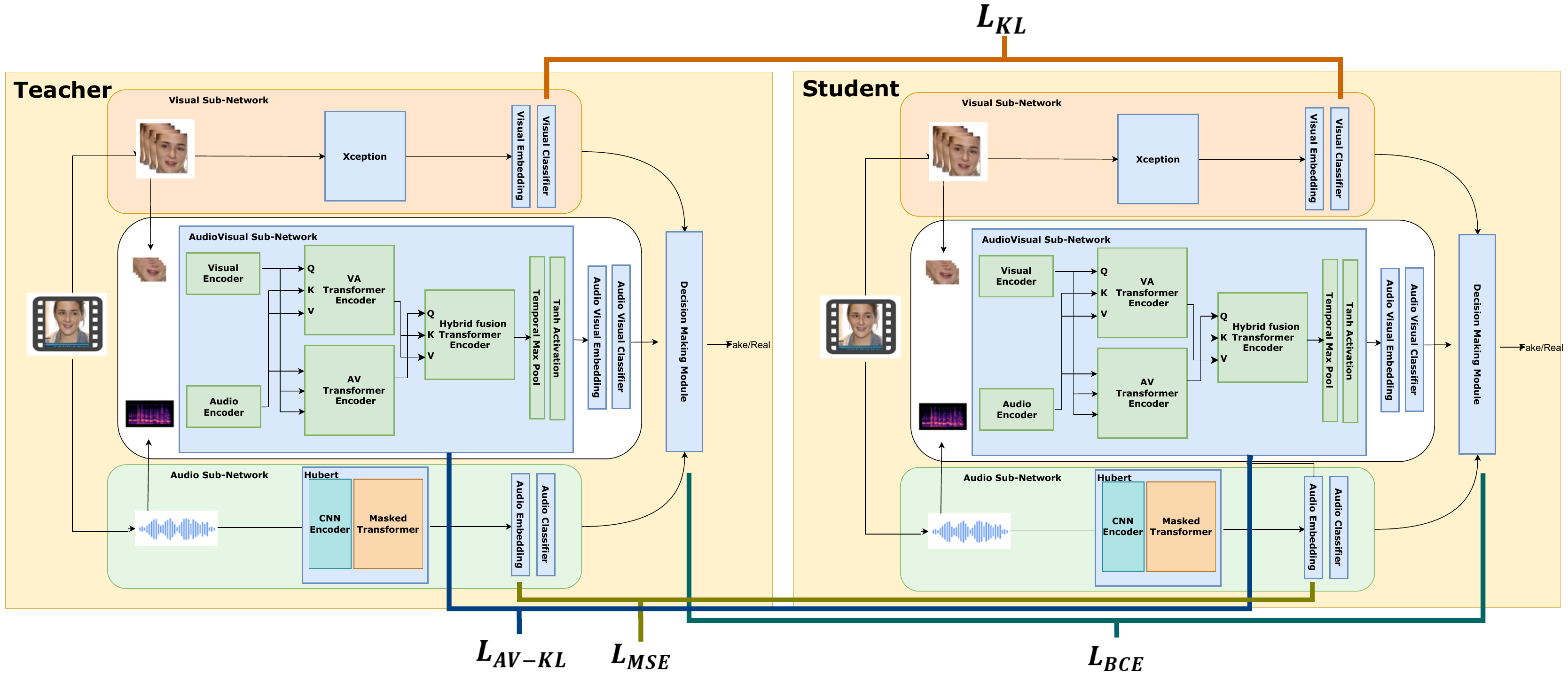}
    \caption{Illustration of our teacher-student framework. The figure depicts the four defined loss functions used to train the student model.}
    \label{fig2}
  \end{center}
\end{figure*}

For this purpose, we first train the teacher model using the loss functions mentioned in the previous section. We then introduce a teacher-student structure, allowing the model to adapt and learn from new data, as illustrated in Figure \ref{fig2}.
In this approach, we choose the same model architecture for both teacher and student models, and also use the weights of the teacher model as the initialization for the student model. Next, we define four loss functions to train the student model on a small subset of unseen and main domain data. This method allows the model to acquire new knowledge guided by the teacher model while preserving its performance on the old domain.

Subsequently, we attempt to train the student model using the teacher's knowledge through specific loss functions. First, we apply a binary cross-entropy loss function to the three sub-networks $f_{S(m)}$.  The term $L_{BCE}$ is equivalent to the first part of the loss function used to train the teacher model, with an additional part that defines this loss function on the final output determined by the decision module.

Then, we apply a mean squared error loss $L_{M S E}$ between the learned features of each sub-network, denoted as $h_\text{S(m)}$ for the student model and $h_\text{T(m)}$ for the teacher model, to enforce feature alignment between them. This regularization encourages the student model to stay close to the feature space of the teacher model.

Since the audio-visual sub-network is the core component of our model, we design specific loss functions to enhance student learning. Inspired by the transformer-based approach in \cite{wang2020minilm}, we apply the KL divergence function between Query-Key and Value-Value matrices of the transformer layers. This ensures the student model remains aligned with the teacher model’s behavior and prevents significant deviations. We define these loss functions in Equation \ref{eq:6}, where $L_\text{CAD}$ reflects the Query-Key term and $L_\text{VR}$ corresponds to the Value-Value term.

\begin{equation} \label{eq:6}
{\small
\begin{gathered}
C A D=\operatorname{softmax}\left(\frac{Q K^T}{\tau \sqrt{d_q}}\right),VR=\operatorname{softmax}\left(\frac{V V^T}{\tau \sqrt{d_v}}\right) \\
L_{C A D}=\sum_{l \in L_t} D_{K L}\left(C A D_{s, l} \| C A D_{t, l}\right) ,  L_{V R}=\sum_{l \in L_t} D_{K L}\left(V R_{s, l} \| V R_{t, l}\right) \\
L_{{AV-KL}}=L_{C A D}+L_{V R} 
\end{gathered}
}
\end{equation}

Similar to other studies \cite{yang2022confidence, wang2020minilm} using the teacher-student framework, we also apply the KL divergence function to the output probabilities of each sub-network to distill class probabilities by minimizing this loss. We also use a dynamic confidence weighting method that adjusts the weights according to the absolute difference between real and fake class probabilities, capturing their evolving importance based on the teacher model.

\begin{equation} \label{eq:7}
{\small
\begin{gathered}
L_{K L}(i, m) = D_{K L}\left(\sigma\left(f_{S(m)}\left(x_{m, i}\right)\right) \| \sigma\left(f_{T(m)}\left(x_{m, i}\right)\right)\right)
\\ \lambda_{i, m}=\left|\left(1-\sigma\left(f_{T(m)}\left(x_{m, i}\right)\right)\right)-\sigma\left(f_{T(m)}\left(x_{m, i}\right)\right)\right| 
\\ L_{K L}=\sum_{m \in\{a, v, a v\}} \frac{1}{N} \sum_{i=1}^N \lambda_{i, m} \times L_{K L}(i, m)
\end{gathered}
}
\end{equation}
Finally, the total student loss is formulated as Equation \ref{eq:8}, where $\beta$, $\gamma$, and $\delta$ are tunable hyperparameters that balance the contributions of the individual loss components, determined through grid search.
% where $\beta$, $\gamma$, and $\delta$ are hyperparameters \newtext
% {controlling the contribution of each loss component, and were tuned through a grid search over a predefined parameter space.}

\begin{equation} \label{eq:8}
{\small
\begin{gathered}
L_{Student}= L_{BCE}+ \beta L_{M S E}+ \gamma L_{AV-KL} + \delta L_{K L}
\end{gathered}
}
\end{equation}
% \begin{figure}[h]
%     \centering
%     \includegraphics[width=\textwidth]{images/P8.pdf}
%     \caption{Proposed Teacher-Student Structure for Domain Adaptation}
%     \label{fig:ts-sturcture}
% \end{figure}

\section{DATASETS AND SETTINGS}
\label{sec:dataset}
\subsection{Datasets}
{
This section outlines the datasets used in our experiments. We selected datasets containing manipulations in audio and/or visual modalities. Since training the teacher model requires explicit labels indicating the modified modality, we used FakeAVCeleb \cite{khalid2021fakeavceleb}, which provides all four label types—RealVideo-RealAudio (RR), RealVideo-FakeAudio (RF), FakeVideo-RealAudio (FR), and FakeVideo-FakeAudio (FF)—necessary for sub-network-specific losses. DeepfakeTIMIT \cite{korshunov2018deepfakes}, DFDC \cite{dolhansky2020deepfake}, and PolyGlotFake \cite{hou2024polyglotfake}, which lack one or more label types, were used exclusively for student training and generalization evaluation.
}

\textbf{FakeAVCeleb} is a multimodal dataset designed for deepfake detection,  focusing on both audio and visual manipulations. It consists of 500 real videos and 19,500 deepfake videos, categorized into four distinct types (RR, RF, FR, FF). The real videos in this dataset are sourced from the VoxCeleb2 \cite{chung2018voxceleb2} corpus that contains videos of celebrities with a wide range of ages and ethnicities, supporting the development of a fair and robust deepfake detection model.

\textbf{DeepfakeTIMIT} constructed using data from 32 selected individuals (forming 16 similar-looking pairs) from the VidTIMIT
\cite{sanderson2009multi}
 database, which contains 10 real video recordings per subject reciting short sentences. Then two face-swapping models (high and low quality) were trained on these videos, generating 620 deepfake videos, while the audio remains unchanged in all cases. In our study, we use these deepfake videos alongside 320 real ones for training the student model and domain adaptation testing.

\textbf{DFDC}
is a large collection of videos created to support research in detecting manipulated media, specifically deepfakes. The dataset includes deepfakes produced using various synthesis techniques.
We randomly select about 6200 videos from the dataset with a 1:3 real/fake ratio. For student model training, we manually relabeled 160 selected videos into four fine-grained categories aligned with the main dataset.
The rest were used for validation and domain adaptation testing.

\newtext{
\textbf{PolyGlotFake}
is a multilingual and multimodal deepfake dataset comprising content in seven languages, generated using advanced text-to-speech, voice cloning, and lip-sync technologies. We select a subset of 3,200 videos, which includes all 766 authentic videos and the remainder as synthetic. A small portion of this subset is used to train the student model, while the rest is reserved for evaluation.
}
\subsection{Implementation Details}
\subsubsection{Pre-processing Pipeline}
To prepare video inputs for each sub-network, we begin by separating audio and frames of video by sampling the audio at 16 kHz and capturing frames at 25 fps. Since our model input is a video clip with C frames, we split each video into smaller clips of frames with their corresponding audio. For audio input, we fed raw sampled audio into the audio sub-network, while we calculated the normalized Mel-spectrogram for the audio input of the audio-visual sub-network. For visual input, we first use the S3FD face detection method
\cite{zhang2017s3fd} to extract faces from the frames. Then, we give the mouth parts of the faces as the visual input to the audio-visual sub-network, and since the input of the visual sub-network is one frame, we randomly select one of these frames of the input clip for this part.
\subsubsection{Training and Evaluating}
In training the teacher model, we randomly select $70\%$ of the main dataset videos for training and the rest for testing. Then we split each video into smaller clips containing C frames. After that, due to the dataset's imbalance, where fake samples dominate, we randomly pick one clip from each fake video while using all clips from each real video. For training, we use the Adam optimizer with a 0.0001 learning rate and the StepLR scheduler. To enhance the model's generalization, we also used different augmentation methods on the input frames, such as brightness modification, rotation, and compression. 

The student model is trained using the same setup on three selected datasets comprising unseen and main domain data, \newtext{gathered manually or via active learning \cite{campbell2000query}}. \textbf{Dataset I} consists of 160 samples from the unseen domain DFDC dataset and  110 additional samples from the main domain dataset FakeAVCeleb. \textbf{Dataset II} includes 80 samples from TIMIT\_Deepfake and 80 from the main dataset, \newtext{while \textbf{Dataset III} uses 100 samples from PolyGlotFake and 80 from the main dataset}. All three datasets are divided into training and validation sets using a 75:25 ratio.\\
Implementation of the proposed model is publicly available \footnote{https://github.com/elhamabolhasani/EAV-DFD}.
\section{EXPERIMENTS AND RESULTS}
This section starts by evaluating a teacher model trained on the main dataset. Using the proposed teacher-student framework, we then train some student models on Dataset I, Dataset II, and \newtext{Dataset III}. These datasets contain a mixture of seen-domain videos (from the main dataset) and unseen-domain videos (from three new datasets). Finally, we evaluate the student model using various loss functions and analyze its performance and explainability. \newtext{All experiments in this section were conducted using an NVIDIA A100 GPU. Average response times and t-SNE embedding plots for both teacher and student models are provided in the Supplementary Material.}

\label{sec:experiments}
\subsection{Teacher Model Evaluation}

\subsubsection{Comparison of Unimodal and Multimodal Approaches}

To demonstrate the superiority of our teacher model and the contribution of the audio and visual sub-networks, we measure the performance of each sub-network individually and in combination using the test data from the main dataset. \newtext{As shown in Table \ref{tab:multimodal_eval}, our model achieves impressive results across audio, visual, and ensemble modes when consistently surpassing other models in both unimodal and multimodal scenarios}. This outcome may be due to the label definition process during training. We define the final label by assuming that each modality can be manipulated, and the network only outputs the real label when all sub-networks agree. This unimodal detection capability also makes the model adaptable to single-modality inputs, enhancing its applicability across various scenarios.

\begin{table}[h]
\caption{Evaluation of teacher model on FakeAVCeleb dataset under different modalities: visual sub-network, audio sub-network, and Ensemble Model (‘–’ in all tables indicates that the corresponding metric was not reported in the original work)}
\label{tab:multimodal_eval}
\centering
\resizebox{\columnwidth}{!}{%
\begin{tabular}{c|c c|c c|c c}
\hline
\multirow{ 2}{*}{\textbf{Model}}&
\multicolumn{2}{c|}{\textbf{\newtext{Visual Sub-Network}}} & \multicolumn{2}{c|}{\textbf{\newtext{Audio Sub-Network}}} & \multicolumn{2}{c}{\textbf{Ensemble Model}} \\ \cline{2-7}
               & \textbf{Accuracy} & \textbf{AUC} & \textbf{Accuracy} & \textbf{AUC} & \textbf{Accuracy} & \textbf{AUC} \\ \hline
               
\textbf{AVA-CL \cite{zhang2023joint}} & 81.76 & 83.87 & 97.50 & 99.85 & 86.55 & 89.47 \\
\textbf{AVTENet \cite{hashmi2025avtenet}} & 85.25 & – & 85.12 & – & 99.00 & – \\
\textbf{AVFakeNet \cite{ilyas2023avfakenet}} & 90.94 & 90.65 & 98.73 & 98.66 & 93.41 & 84.67 \\
\textbf{AVT²–DWF \cite{10609529}} & 70.95 & – & – & – & 87.57 & 88.32 \\

\newtext{\textbf{STKD-VViT} \cite{usmani2025spatio}} & \newtext{97.49}  & \newtext{–} & \newtext{98.65} & \newtext{-} & \newtext{96.0} & \newtext{-} \\

\newtext{\textbf{SS-AVD} \cite{zhang2025lightweight}} & \newtext{98.11} & \newtext{99.37} &  \newtext{98.55} & \newtext{99.51} &  \newtext{95.11} & \newtext{98.31} \\

\hline
\textbf{EAV-DFD} & \textbf{99.09} & \textbf{99.64} & \textbf{99.65} & \textbf{99.99} & \textbf{99.33} & \textbf{99.88} \\ \hline
\end{tabular}%
}
\end{table}

\subsubsection{Cross-Dataset Generalization}

In this part, we show the generalization ability of our teacher model by evaluating it on three cross-datasets. The results, which are detailed in Table \ref{tab:cross_dataset_eval}, indicate that the model performs effectively on these unseen datasets. However, as highlighted in the introduction, achieving better results on new datasets requires more than just generalization methods, and adaptation techniques are essential to handle the domain shift in new data and improve performance. 

\begin{table}[h]
\caption{Cross-dataset evaluation of teacher model on FakeAVCeleb and three unseen datasets}
\label{tab:cross_dataset_eval}
\centering
\resizebox{\columnwidth}{!}{%
\begin{tabular}{c|c|c c|c c|c c|c c}
\hline
\multirow{ 3}{*}{\textbf{Model}}&
\multirow{ 3}{*}{\textbf{\#Clips}}&
\multicolumn{6}{c}{\textbf{Trained on FakeAVCeleb}}\\ \cline{3-10}
 & &\multicolumn{2}{c|}{\textbf{FakeAVCeleb}}&
 \multicolumn{2}{c|}{\textbf{TIMIT\_Deepfake}} & \multicolumn{2}{c|}{\textbf{DFDC}} &
 \multicolumn{2}{c}{\textbf{\newtext{PolyGoltFake}}}\\ \cline{3-10}
               &        & \textbf{Accuracy} & \textbf{AUC} & \textbf{Accuracy} & \textbf{AUC} & \textbf{Accuracy} & \textbf{AUC} & \textbf{\newtext{Accuracy}} & \newtext{\textbf{AUC}}\\ \hline
\multirow{ 4}{*}{\textbf{Teacher Model}}   & 1 & 99.26 & 98.49  & 71.41 & 74.73 & 71.40 & 60.23 & \newtext{90.04} & \newtext{86.05}\\
& 3    & 99.39 & 99.85    & \textbf{72.98} & 79.27 & 75.41 & 65.01 & \newtext{92.04} & \newtext{94.81}\\
& 7   & 99.33 & \textbf{99.88}   & 72.25 & \textbf{81.71} & 76.11 & 67.33 & \newtext{93.25} & \newtext{97.27} \\ 
  & 9   & \textbf{99.45} & 99.92    & 72.77 & 81.47 & \textbf{76.29} & \textbf{69.00} & \newtext{\textbf{93.37}} & \newtext{\textbf{97.76}}\\ \hline
\end{tabular}%
}
\end{table}

\begin{table*}[ht!]
\centering
\caption{Performance comparison of different models across four datasets when the model is trained on FakeAVCeleb. \text{‡}: the model is reproduced by \cite{10609529} and the authors did not report accuracy/auc metrics on these datasets}
\resizebox{0.65\textwidth}{!}{%
\begin{tabular}{c|cc|cc|cc|cc}
\hline
\multirow{ 2}{*}{\textbf{Model}}&
\multicolumn{2}{c|}{\textbf{FakeAVCeleb}} & \multicolumn{2}{c|}{\textbf{TIMIT\_Deepfake}} & \multicolumn{2}{c|}{\textbf{DFDC}} &
\multicolumn{2}{c}{\textbf{\newtext{PolyGlotFake}}} \\ \cline{2-9}
               & \textbf{Accuracy} & \textbf{AUC} & \textbf{Accuracy} & \textbf{AUC} & \textbf{Accuracy} & \textbf{AUC} & \textbf{\newtext{Accuracy}} & \textbf{\newtext{AUC}} \\ \hline
               
\textbf{Xception} \cite{rossler2019faceforensics++}‡       & 72.71 & 73.51 & - & 64.04 & - & 50 & \newtext{-} & \newtext{-} \\
\textbf{MDS} \cite{chugh2020not}‡       &  81.80 & 82.65 & - & 53 & - & 63 & \newtext{-} & \newtext{-} \\
\textbf{LipForensics} \cite{haliassos2021lips}‡        & 64.00 &  65.23  & - & 55.27 & - & 49 & \newtext{-} & \newtext{-}\\

\textbf{AVA-CL} \cite{zhang2023joint} & 86.55 & 89.47 & -    & 88.34  & -    & 65.88 & \newtext{-} & \newtext{-}  \\
\textbf{AVT$^2$-DWF} \cite{10609529}        & 87.57 & 88.32 & -    & 67.50 & -    & \textbf{74.60} & \newtext{-} & \newtext{-} \\ \hline
\textbf{EAV-DFD}              & \textbf{99.33} & \textbf{99.88} & 72.25 & 81.71 & 76.11 & 67.33  & \newtext{93.25} & \newtext{97.27} \\
\textbf{EAV-DFD Student}  & 98.13 & 99.02 & \textbf{97.02} & \textbf{99.65} & \textbf{77.67} & 71.42 & \newtext{\textbf{95.76}} & \newtext{\textbf{97.77}} \\
\hline
\end{tabular}%
}
\label{tab:model_comparison}
\end{table*}

\subsection{Student Model Evaluation}
After training the teacher model, we proceed to train a student model using our proposed teacher-student framework. Thus, we train the student model using Dataset I \newtext{($\beta= \gamma=\delta=2$)}, Dataset II, and Dataset III ($\beta= \gamma=\delta=4$), which consist of a small number of videos from the main and new unseen datasets. \newtext{The hyperparameters were tuned using a grid search}. As shown in Table \ref{tab:stu}, the student model exhibits a slight decrease in performance on the main dataset, while improving the AUC of the student model by 4.09\%, 17.94\%, and 0.5\% on the new datasets compared to the teacher model. This showcases the effectiveness of our approach in enhancing models using this framework when dealing with newly invented deepfake generation methods. \newtext{(Additional results for each sub-network performance for Dataset III are provided in the Supplementary Material (Table II, III))}

\begin{table}[h]
\caption{Comparison of student and teacher model performance after training the student model on three student datasets}
\label{tab:stu}
\centering
\resizebox{0.7\columnwidth}{!}{%
\begin{tabular}{c|c c|c c}
\hline
\multirow{3}{*}{\textbf{Model}}&
\multicolumn{4}{c}{\textbf{Train Data (Dataset I)}}  \\ \cline{2-5} & \multicolumn{2}{c|}{\textbf{DFDC}} & \multicolumn{2}{c}{\textbf{FakeAVCeleb}} \\ \cline{2-5}
& \textbf{Accuracy} & \textbf{AUC} & \textbf{Accuracy} & \textbf{AUC} \\ \hline
\textbf{Teacher Model}     & 76.11 & 67.33 & \textbf{99.33} & \textbf{99.88} \\
\textbf{Student Model}  & \textbf{77.67} & \textbf{71.42} &97.87 & 97.87  \\ \hline
\end{tabular}%
}
\\[0.1cm]

\resizebox{0.7\columnwidth}{!}{%
\begin{tabular}{c|c c|c c}
\hline
\multirow{3}{*}{\textbf{Model}}&
\multicolumn{4}{|c}{\textbf{Train Data (Dataset II)}} \\ \cline{2-5} &  \multicolumn{2}{|c|}{\textbf{TIMIT\_Deepfake}} & \multicolumn{2}{c}{\textbf{FakeAVCeleb}} \\ \cline{2-5}
& \textbf{Accuracy} & \textbf{AUC} & \textbf{Accuracy} & \textbf{AUC} \\ \hline
\textbf{Teacher Model}     & 72.25 & 81.71 & \textbf{99.33} & \textbf{99.88} \\
\textbf{Student Model}  & \textbf{97.02} & \textbf{99.65} & 98.13 & 99.02 \\ \hline
\end{tabular}%
}
\\[0.1cm]
\resizebox{0.7\columnwidth}{!}{%
\color{reviewcolor}
\begin{tabular}{c|c c|c c}
\hline
\multirow{3}{*}{\textbf{Model}}&
\multicolumn{4}{|c}{\textbf{Train Data (Dataset III)}} \\ \cline{2-5} &  \multicolumn{2}{|c|}{\textbf{PloyGlotFake}} & \multicolumn{2}{c}{\textbf{FakeAVCeleb}} \\ \cline{2-5}
& \textbf{Accuracy} & \textbf{AUC} & \textbf{Accuracy} & \textbf{AUC} \\ \hline
\textbf{Teacher Model}     & 93.25 & 97.27 & 99.33 & 99.88 \\
\textbf{Student Model}  & \textbf{95.76} & \textbf{97.77} & \textbf{99.81} & \textbf{99.94}  \\ \hline
\end{tabular}%
}

\end{table}

We then compare the results of our models with those of other existing deepfake detection models. As shown in Table \ref{tab:model_comparison}, the teacher model outperforms other state-of-the-art models on the main dataset and demonstrates good cross-dataset generalization.
Student models also, by employing the teacher-student framework, achieve better cross-dataset performance while utilizing a limited number of training videos from the unseen datasets. On the Deepfake\_TIMIT dataset, the student model achieved strong results with a significant margin, demonstrating the robustness of our approach. \newtext{For the PolyGlotFake dataset, the student model attained higher accuracy, while the teacher model also performed well on this dataset}. On the DFDC dataset, while the teacher model outperforms AVT²–DWF in both unimodal and multimodal evaluations on the main dataset (as mentioned in Table \ref{tab:multimodal_eval}), the student model does not achieve better AUC results. However, due to the dataset's complexity, the student model may need more data and training to reach a higher AUC metric.

\subsection{{Ablation Study}}
\subsubsection{{Teacher Model Fusion Strategies}}
{We evaluate three strategies for fusing the sub-networks of the teacher model: (i) Split, where each sub-network has its own classifier and the final decision is made according to Algorithm 1 in the paper; (ii) Joint, where embeddings are concatenated and passed to a single classifier; and (iii) Attention, where an attention layer is applied over the three embeddings to model inter-modality correlations. Table \ref{tab:fusion_strategies} presents the corresponding evaluation results. As shown, although the Joint model achieved slightly higher accuracy on the FakeAVCeleb dataset, the Split configuration consistently outperformed both the Joint and Attention models in cross-dataset evaluations—showing superior generalization across two of the three cross-domain benchmarks. Therefore, we selected the Split model for applying the teacher–student framework due to its stronger generalization ability.}

 \begin{table}[ht]
    \centering
    
    \caption{Performance comparison of models with different fusion strategies across four datasets when the model is trained on FakeAVCeleb}
    \resizebox{\columnwidth}{!}{%
    \begin{tabular}{c|c|c|c|c|c|c|c|c}
    \hline
    \multirow{3}{*}{Model} & \multicolumn{2}{c|}{FakeAVCeleb} & \multicolumn{2}{c|}{DFDC} & \multicolumn{2}{c|}{TIMIT\_Deepfake} & \multicolumn{2}{c}{PloyGlotFake} \\
    \cline{2-9}
     & Accuracy & AUC & Accuracy & AUC & Accuracy & AUC & Accuracy & AUC \\
    \hline
    Teacher (Split) & 99.33 & 99.88 & \textbf{76.11} & \textbf{67.33} & 72.25 & 81.71 & \textbf{93.25} & \textbf{97.27} \\
    \hline
    Teacher (Joint) & \textbf{99.81} & \textbf{99.94} & 39.93 & 63.19 & 65.91 & 65.91 & 69.75 & 89.61 \\
    \hline
    Teacher (Attention) & 98.66 & 99.81 & 60.47 & 42.55 & \textbf{77.30} & \textbf{88.42} & 66.39 & 86.86 \\
    \hline
    \end{tabular}%
    }
    \label{tab:fusion_strategies}
    \end{table}

\subsubsection{{Impact of Frame Count per Clip ($C$)}}
{
After that, we analyze the effect of video clip size on the teacher model's performance. So, we train a model using different values for hyperparameter C as reported in Table \ref{tab:differentclipsizes}, where the best and second-best values are highlighted in bold and underlined, respectively. Then we choose an input size of 20 for this hyperparameter based on the better performance of the model on the unseen datasets.
}

\begin{table}[ht]
\centering
\caption{Evaluation of \newtext{teacher} model with different clip sizes}
\resizebox{\columnwidth}{!}{%
\begin{tabular}{c|c|c c|c c|c c| c c}
\hline
\multirow{ 3}{*}{\textbf{Frame Numbers (C)}}&
\multirow{ 3}{*}{\textbf{\#Clips}}& \multicolumn{6}{c}{\textbf{Test Datasets}}\\ \cline{3-10}
& & \multicolumn{2}{c|}{\textbf{FakeAVCeleb}} & \multicolumn{2}{c|}{\textbf{DFDC}} & \multicolumn{2}{c|}{\textbf{TIMIT\_Deepfake}} &
\multicolumn{2}{c}{\textbf{\newtext{PolyGlotFake}}} \\ \cline{3-10}
& &\textbf{Accuracy} & \textbf{AUC} & \textbf{Accuracy} & \textbf{AUC} & \textbf{Accuracy} & \textbf{AUC} & \textbf{\newtext{Accuracy}} & \textbf{\newtext{AUC}} \\ \hline
\textbf{10} & 14 & \underline{99.61} & 99.75 & 67.30 & \textbf{71.62} & \underline{69.63} & \underline{80.23} & \newtext{\textbf{93.53}} & \newtext{\textbf{98.55}} \\ 
\textbf{20} & 7  & 99.33 & \underline{99.88} & \textbf{76.11} & \underline{67.33} & \textbf{72.25} & \textbf{81.71} & \newtext{\underline{93.25}} & \newtext{\underline{97.27}} \\ 
\textbf{30} & 5  & \textbf{99.76} & \textbf{99.95} & \underline{71.38} & 62.73 & 68.27 & 72.97 & \newtext{90.47} & \newtext{96.68} \\ \hline
\end{tabular}%
}
\label{tab:differentclipsizes}
\end{table}

\subsubsection{{Impact of Video Clips Numbers per Sample ($N$)}}
{Building on the results reported in Table~\ref{tab:cross_dataset_eval}, we examine how the number of video clips per sample affects the final prediction. As expected, increasing the number of clips leads to performance gains; however, after a certain threshold, the improvement becomes negligible, and the additional computational cost is not justified.  Therefore, we set the number of clips to 7 for all subsequent experiments.}

\subsubsection{{Evaluation of Student Loss Functions}}
{To investigate optimization behavior, we also train the student model with different combinations of loss functions. Table \ref{tab:loss_performance} illustrates how each loss function term contributes to improving performance. Notably, employing only the $L_{BCE}$ term in the loss function yields a significant improvement, increasing AUC by 17.87\% on the Deepfake\_TIMIT dataset.
While adding two other $L_{AV-KL}$ and $L_{MSE}$ terms to the loss function does not further boost AUC, it enhances accuracy. The results also confirm that dynamic weighting further refines model performance.}

\begin{table}
\centering
\caption{Performance of the student model trained on Dataset II using different loss functions}
\resizebox{\columnwidth}{!}{%
\begin{tabular}{c|c c c c c|c c|c c}
    \hline
    \multirow{3}{*}\textbf{Model} & \multicolumn{5}{c|}{\textbf{Loss Function}} & \multicolumn{2}{c|}{\textbf{FakeAVCeleb}} & \multicolumn{2}{c}{\textbf{TIMIT\_Deepfake}} \\ \cline{2-10}
    & \textbf{BCE} & \textbf{AV-KL} & \textbf{MSE} & \textbf{KL-NW} & \textbf{KL-W} & \textbf{Accuracy} & \textbf{AUC} & \textbf{Accuracy} & \textbf{AUC} \\ \hline
    \textbf{Teacher model} & - & - & - & - & - & \textbf{99.33} & \textbf{99.88} & 72.25 & 81.71 \\ \hline
    \multirow{4}{*}{\textbf{Student model}} 
    & \checkmark & - & - & - & - & 99.31 & 99.79 & 83.27 & 99.58 \\ \cline{2-10}
    & \checkmark & \checkmark & \checkmark & - & - & 98.15 & 96.82 & 93.39 & 98.60 \\  \cline{2-10}
    & \checkmark & \checkmark & \checkmark & \checkmark & - & 98.29 & 99.23 & 95.59 & 99.29 \\  \cline{2-10}
    & \checkmark & \checkmark & \checkmark & - & \checkmark & 98.13 & 99.02 & \textbf{97.02} & \textbf{99.65} \\  \hline
    \bottomrule
\end{tabular}%
}
\label{tab:loss_performance}
\end{table}

\subsubsection{{Effect of Student Dataset size}}
\begin{figure}[h]
    \centering
    \color{reviewcolor}
    \includegraphics[width=0.95\columnwidth]{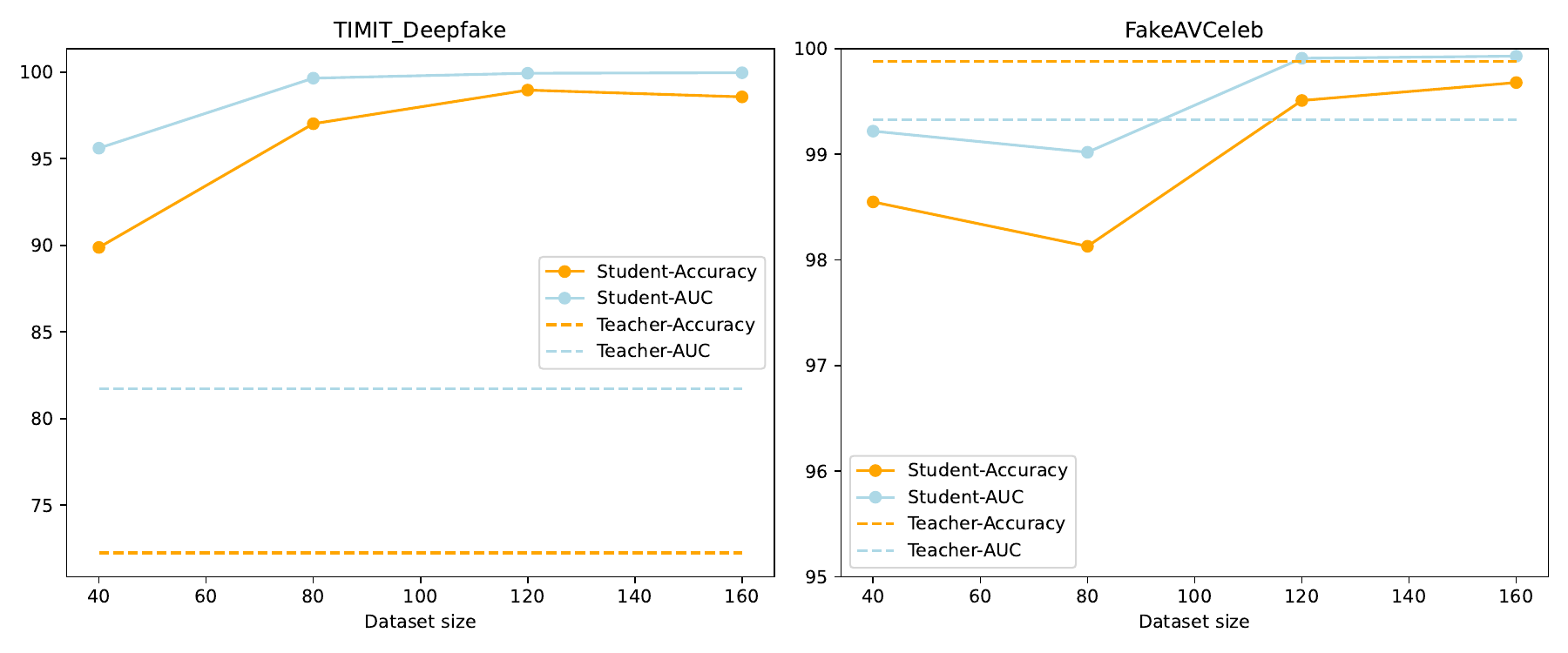}
    \caption{
    Comparison of student and teacher model performance (Accuracy and AUC) across varying dataset sizes (number of samples per dataset) for the TIMIT\_Deepfake and FakeAVCeleb datasets.}
    \label{different_dataset_size}
\end{figure}

{We analyze the effect of varying the size of the student training dataset on model performance to show our approach’s scalability. As seen in Figure  \ref{different_dataset_size}, performance improves as the dataset grows. However, adding over 120 samples per dataset offers little benefit. This trend shows a saturation point; gathering more data becomes time and cost-inefficient with minor performance gains. Because of our low-data focus, smaller and optimized datasets boost resource efficiency and model effectiveness. For this reason, increasing the sample size beyond this point is not recommended because of the minimal advantages compared to the added effort.
}

\subsection{\newtext{Modality-Level Explainability}}
\begin{figure}[h]
    \centering
    \includegraphics[width=0.8\columnwidth,height=3cm]{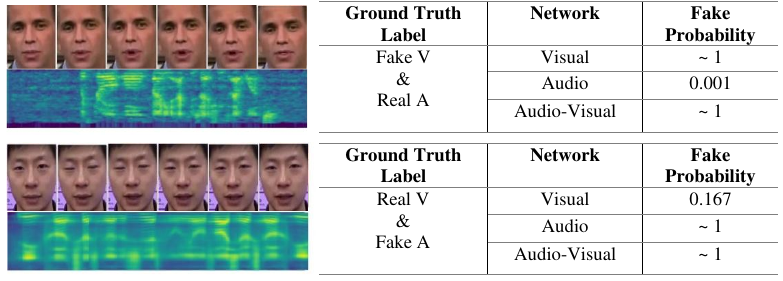}
    \caption{Output probabilities of each sub-network for two fake samples with FakeVideo-RealAudio and RealVideo-FakeAudio labels.}
    \label{fig3}
\end{figure}

In this study, we introduce a model designed as an ensemble of audio-visual and unimodal sub-networks. The training process of our model also ensures that each of the three sub-networks' functions can perform well independently. This design allows us to analyze the output probabilities and identify manipulated modalities in the video inputs. Figure \ref{fig3} shows the output probabilities for two samples with different labels.
For example, in a sample labeled FakeVideo-RealAudio, the audio-visual and visual sub-networks detect the video as fake with high probability. At the same time, the audio sub-network identifies the audio as real. These probabilities suggest that manipulation occurred within the video frames. Given that the audio-visual sub-network, which examines the correlation between lip movements and speech, also detects the video as fake, we can infer that the manipulation likely targets the mouth region. This interpretation aligns with the face-swap deepfake generation method used to produce this sample, demonstrating the explainability of our approach.

\newtext{
\subsection{Model Robustness Analysis}
In this section, we assess the robustness of our student model under challenging conditions such as noise, compression, and cropping on both modalities. During the training of both the teacher and the student models, we employ augmentation techniques —Gaussian noise, blur, rotation, frame compression, and brightness alteration— to improve the robustness of our model. Table \ref{tab:differentnoise} presents the performance of the student and teacher models under various perturbations. On the main FakeAVCeleb dataset, both models maintain strong performance with only a slight decrease. Notably, on the cross TIMIT\_Deepfake dataset, the student model demonstrates greater robustness, highlighting the effectiveness of our approach.
}

%(\mu=0, \sigma=0.05)
\begin{table}[ht]
\centering
\color{reviewcolor}
\caption{
Performance Comparison of Teacher and Student (Dataset II)  Models under Perturbations on FakeAVCeleb and TIMIT\_Deepfake }
%Performance Comparison of Teacher and Student Models Under Different Perturbations
\resizebox{0.8\columnwidth}{!}{%
\begin{tabular}{c|c|c|c|c|c}
\hline
\multirow{2}{*}{\textbf{Perturbation Type}} & \multirow{2}{*}{\textbf{Model}} & \multicolumn{2}{c|}{\textbf{FakeAVCeleb}} & \multicolumn{2}{c}{\textbf{TIMIT\_Deepfake}} \\
\cline{3-6}
 & & \textbf{Accuracy} & \textbf{AUC} & \textbf{Accuracy} & \textbf{AUC} \\
\hline
\multirow{3}{*}{\textbf{Without Noise}} & Teacher & 99.33 & 99.88 & 72.25 & 81.71\\
 & Student& 98.12 & 99.02 & 97.02 & 99.65\\
\hline
\multirow{3}{*}{\textbf{Gaussian Noise}} & Teacher & 98.19 & 99.16 & 70.04 & 66.57 \\
 & Student& 98.21 & 93.49 & 70.68 & 79.25\\
\hline
\multirow{3}{*}{\textbf{Compressed}} & Teacher & 93.53 & 97.57 & 72.50 & 83.06\\
 & Student& 97.70 & 63.13 & 90.01 & 98.24\\
\hline
\multirow{3}{*}{\textbf{Cropped}} & Teacher & 99.28 & 99.94 & 73.28 & 81.35\\
 & Student& 98.26 & 98.77 & 94.94 & 98.82 \\
\hline
\end{tabular}

}
\label{tab:differentnoise}
\end{table}
% \begin{table}[ht]
% \centering
% \color{reviewcolor}
% \caption{Performance Comparison of Teacher and Student Models Under Different Perturbations}
% \resizebox{\columnwidth}{!}{%
% \begin{tabular}{c|c|c|c|c|c}
% \hline
% \multirow{2}{*}{\textbf{Perturbation Type}} & \multirow{2}{*}{\textbf{Model}} & \multicolumn{2}{c|}{\textbf{FakeAVCeleb}} & \multicolumn{2}{c}{\textbf{TIMIT\_Deepfake}} \\
% \cline{3-6}
%  & & \textbf{Accuracy} & \textbf{AUC} & \textbf{Accuracy} & \textbf{AUC} \\
% \hline
% \multirow{3}{*}{Without Noise} & Teacher & 99.33 & 99.88 & 72.25 & 81.71\\
%  & Student (Dataset II) & 98.12 & 99.02 & 97.02 & 99.65\\
% \hline
% \multirow{3}{*}{Gaussian Noise (\mu=0, \sigma=0.05)} & Teacher & 98.19 & 99.16 & 70.04 & 66.57 \\
%  & Student (Dataset II) & 98.21 & 93.49 & 70.68 & 79.25\\
% \hline
% \multirow{3}{*}{Compressed} & Teacher & 93.53 & 97.57 & 72.50 & 83.06\\
%  & Student (Dataset II) & 97.70 & 63.13 & 90.01 & 98.24\\
% \hline
% \multirow{3}{*}{Cropped} & Teacher & 99.28 & 99.94 & 73.28 & 81.35\\
%  & Student (Dataset II) & 98.26 & 98.77 & 94.94 & 98.82 \\
% \hline
% \end{tabular}

\begin{figure}[h]
    \centering
    \includegraphics[width=0.8\columnwidth]{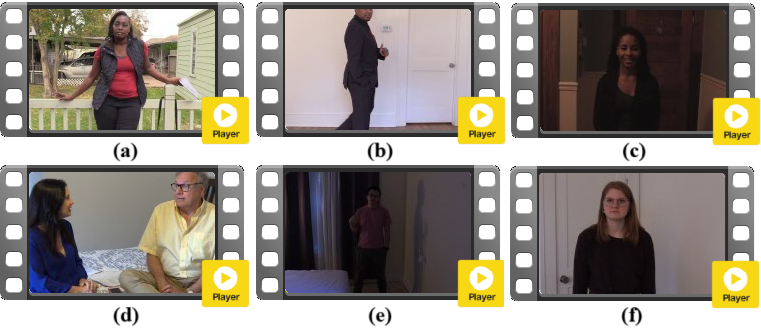}
    \caption{
    \newtext{
    Examples of false negatives (top) and false positives (bottom) produced by EAV-DFD. (a) fake without visual or audio artifacts; (b) face occlusion due to motion. (c) and (e) low-light cases. (d) multiple speakers, and (f) an off-screen speaker prompting a facial reaction, leading to confusion.
    }
    }
    \label{fig-error}
\end{figure}

% }
% \label{tab:differentnoise}
% \end{table}
\newtext{
\subsection{Failure Cases and Error Analysis}
% To better understand the limitations of our approach, we conducted a detailed error analysis.
% Here, we analyze failure cases to characterize the limitations of the proposed method better. Our model is designed for single-speaker videos and performs well on this task.
% The FakeAVCeleb, TIMIT\_Deepfake, and PolyGlotFake datasets used in our model primarily consist of videos featuring a single person speaking —a scenario for which our model is specifically designed and performs well. 
% However, the DFDC dataset presents a greater challenge due to its more complex video content. 

Here, we focus on analyzing the model's failure cases and discuss potential improvements, such as incorporating additional pre-processing modules to handle such inputs better on DFDC dataset which presents a greater challenge due to its more complex video content. 
In our analysis, several failure modes were identified. First, failures occurred when videos exhibited challenging conditions such as poor lighting, low resolution, or motion-induced distortions, including frame stretching and blurring (Figure \ref{fig-error} (b, c, e)). Second, 
failure cases arose in videos containing multiple faces, multiple speakers, or environmental noise (Figure \ref{fig-error} (d, f)), because our model is designed for single-speaker videos. For example, the model might incorrectly associate the face of one individual with the voice of another, leading the audio-visual network to detect lip-sync mismatches even in authentic videos falsely. Third, in cases where no clues of manipulated audio or visual content are present (Figure \ref{fig-error} (a)).

For future work, depending on the target application, the first issue may be mitigated by applying image enhancement techniques \cite{8449842, wang2025adaptive, zhang2022deep}, or pre-processing techniques to better reveal visual artifacts. The second issue can be addressed through methods such as speaker separation and multi-face tracking to identify and resolve such conflicts. Subsequently, affected sub-networks can be bypassed or assigned reduced weight to prevent error propagation in the final decision. Also, changing the number of clips analyzed per video or the decision threshold based on validation data can improve the model's performance in real-world complex scenarios.
% Our model comprises three specialized sub-networks, each designed to detect a specific type of manipulated content. To ensure robust performance in real-world scenarios, it is essential to calibrate the relative weights of each sub-network, the number of clips analyzed per video, and the decision threshold using validation data.
}

\section{Conclusion}
\label{sec:conclusion}
In this paper, we aimed to create a flexible model that can keep learning from new data without the catastrophic forgetting issue. In this regard, we construct a generalized teacher model and subsequently enhance its efficacy by employing a teacher-student framework for domain adaptation.  Our model incorporates an ensemble architecture that contains audio-visual and unimodal sub-networks, allowing it to process unimodal inputs while providing explainability. Experimental results indicate that the proposed teacher model achieves superior performance compared to other multimodal deepfake detection methods with unimodal detection ability. Our student model also surpasses many deepfake detection methods, which highlights the effectiveness of our proposed teacher-student approach. 
Even though bigger datasets improve models, we tested the model’s efficiency in tight spots using few videos from new areas.
%Although larger datasets can enhance student model performance, our objective was to evaluate the model’s effectiveness under constrained conditions, using only a limited number of video samples from new, unseen domains. 
\newtext{Future work could involve using larger and combined datasets, exploring alternative domain adaptation techniques (e.g., modality-specific adapters in the teacher model’s transformers), and extending explainability toward fine-grained interpretations. It would also be valuable to investigate model distillation to reduce the the audio–visual sub-network, enhance robustness against adversarial attacks, and evaluate the approach on diffusion-based deepfakes once suitable benchmarks become available.}
% Future work could involve using larger and combined datasets, exploring alternative domain adaptation techniques (e.g., modality-specific adapters in the teacher model’s transformers), and extending explainability toward more fine-grained interpretations. It would also be valuable to investigate model distillation to reduce the the audio–visual sub-network, enhance robustness against adversarial attacks, and evaluate the approach on diffusion-based deepfakes once suitable benchmarks become available.

% In future work, we intend to utilize larger datasets, integrate multiple datasets, and investigate other domain adaptation techniques—such as employing adapters in the teacher model’s transformers, where each adapter is specialized for a specific type of deepfake generation method.

\bibliographystyle{IEEEtran}
\bibliography{references}

@inproceedings{yang2019exposing,
  title={Exposing deep fakes using inconsistent head poses},
  author={Yang, Xin and Li, Yuezun and Lyu, Siwei},
  booktitle={ICASSP 2019-2019 IEEE International Conference on Acoustics, Speech and Signal Processing (ICASSP)},
  pages={8261--8265},
  year={2019},
  organization={IEEE}
}

@article{caldelli2021optical,
  title={Optical Flow based CNN for detection of unlearnt deepfake manipulations},
  author={Caldelli, Roberto and Galteri, Leonardo and Amerini, Irene and Del Bimbo, Alberto},
  journal={Pattern Recognition Letters},
  volume={146},
  pages={31--37},
  year={2021},
  publisher={Elsevier}
}

@inproceedings{singh2021detection,
  title={Detection of ai-synthesized speech using cepstral \& bispectral statistics},
  author={Singh, Arun Kumar and Singh, Priyanka},
  booktitle={2021 IEEE 4th International Conference on Multimedia Information Processing and Retrieval (MIPR)},
  pages={412--417},
  year={2021},
  organization={IEEE}
}

@inproceedings{nguyen2019capsule,
  title={Capsule-forensics: Using capsule networks to detect forged images and videos},
  author={Nguyen, Huy H and Yamagishi, Junichi and Echizen, Isao},
  booktitle={ICASSP 2019-2019 IEEE International Conference on Acoustics, Speech and Signal Processing (ICASSP)},
  pages={2307--2311},
  year={2019},
  organization={IEEE}
}

@inproceedings{wang2022m2tr,
  title={M2tr: Multi-modal multi-scale transformers for deepfake detection},
  author={Wang, Junke and Wu, Zuxuan and Ouyang, Wenhao and Han, Xintong and Chen, Jingjing and Jiang, Yu-Gang and Li, Ser-Nam},
  booktitle={Proceedings of the 2022 international conference on multimedia retrieval},
  pages={615--623},
  year={2022}
}

@inproceedings{guera2018deepfake,
  title={Deepfake video detection using recurrent neural networks},
  author={G{\"u}era, David and Delp, Edward J},
  booktitle={2018 15th IEEE international conference on advanced video and signal based surveillance (AVSS)},
  pages={1--6},
  year={2018},
  organization={IEEE}
}

@inproceedings{boccignone2022deepfakes,
  title={Deepfakes have no heart: A simple rppg-based method to reveal fake videos},
  author={Boccignone, Giuseppe and Bursic, Sathya and Cuculo, Vittorio and D’Amelio, Alessandro and Grossi, Giuliano and Lanzarotti, Raffaella and Patania, Sabrina},
  booktitle={International Conference on Image Analysis and Processing},
  pages={186--195},
  year={2022},
  organization={Springer}
}

@inproceedings{haliassos2021lips,
  title={Lips don't lie: A generalisable and robust approach to face forgery detection},
  author={Haliassos, Alexandros and Vougioukas, Konstantinos and Petridis, Stavros and Pantic, Maja},
  booktitle={Proceedings of the IEEE/CVF conference on computer vision and pattern recognition},
  pages={5039--5049},
  year={2021}
}

@inproceedings{conti2022deepfake,
  title={Deepfake speech detection through emotion recognition: a semantic approach},
  author={Conti, Emanuele and Salvi, Davide and Borrelli, Clara and Hosler, Brian and Bestagini, Paolo and Antonacci, Fabio and Sarti, Augusto and Stamm, Matthew C and Tubaro, Stefano},
  booktitle={ICASSP 2022-2022 IEEE International Conference on Acoustics, Speech and Signal Processing (ICASSP)},
  pages={8962--8966},
  year={2022},
  organization={IEEE}
}

@inproceedings{zhou2021joint,
  title={Joint audio-visual deepfake detection},
  author={Zhou, Yipin and Lim, Ser-Nam},
  booktitle={Proceedings of the IEEE/CVF International Conference on Computer Vision},
  pages={14800--14809},
  year={2021}
}

@inproceedings{oorloff2024avff,
  title={AVFF: Audio-Visual Feature Fusion for Video Deepfake Detection},
  author={Oorloff, Trevine and Koppisetti, Surya and Bonettini, Nicol{\`o} and Solanki, Divyaraj and Colman, Ben and Yacoob, Yaser and Shahriyari, Ali and Bharaj, Gaurav},
  booktitle={Proceedings of the IEEE/CVF Conference on Computer Vision and Pattern Recognition},
  pages={27102--27112},
  year={2024}
}

@article{yang2023avoid,
  title={AVoiD-DF: Audio-Visual Joint Learning for Detecting Deepfake},
  author={Yang, Wenyuan and Zhou, Xiaoyu and Chen, Zhikai and Guo, Bofei and Ba, Zhongjie and Xia, Zhihua and Cao, Xiaochun and Ren, Kui},
  journal={IEEE Transactions on Information Forensics and Security},
  volume={18},
  pages={2015--2029},
  year={2023},
  publisher={IEEE}
}

@article{yu2023pvass,
  title={PVASS-MDD: Predictive Visual-audio Alignment Self-supervision for Multimodal Deepfake Detection},
  author={Yu, Yang and Liu, Xiaolong and Ni, Rongrong and Yang, Siyuan and Zhao, Yao and Kot, Alex C},
  journal={IEEE Transactions on Circuits and Systems for Video Technology},
  year={2023},
  publisher={IEEE}
}

@article{liu2023magnifying,
  title={Magnifying multimodal forgery clues for Deepfake detection},
  author={Liu, Xiaolong and Yu, Yang and Li, Xiaolong and Zhao, Yao},
  journal={Signal Processing: Image Communication},
  volume={118},
  pages={117010},
  year={2023},
  publisher={Elsevier}
}

@inproceedings{lee2021tar,
  title={Tar: Generalized forensic framework to detect deepfakes using weakly supervised learning},
  author={Lee, Sangyup and Tariq, Shahroz and Kim, Junyaup and Woo, Simon S},
  booktitle={IFIP International Conference on ICT Systems Security and Privacy Protection},
  pages={351--366},
  year={2021},
  organization={Springer}
}

@inproceedings{khan2021video,
  title={Video transformer for deepfake detection with incremental learning},
  author={Khan, Sohail Ahmed and Dai, Hang},
  booktitle={Proceedings of the 29th ACM International Conference on Multimedia},
  pages={1821--1828},
  year={2021}
}

@inproceedings{kim2021fretal,
  title={Fretal: Generalizing deepfake detection using knowledge distillation and representation learning},
  author={Kim, Minha and Tariq, Shahroz and Woo, Simon S},
  booktitle={Proceedings of the IEEE/CVF conference on computer vision and pattern recognition},
  pages={1001--1012},
  year={2021}
}

@article{kadandale2022vocalist,
  title={Vocalist: An audio-visual synchronisation model for lips and voices},
  author={Kadandale, Venkatesh S and Montesinos, Juan F and Haro, Gloria},
  journal={arXiv preprint arXiv:2204.02090},
  year={2022}
}

@inproceedings{yang2022confidence,
  title={Confidence-Calibrated Face Image Forgery Detection with Contrastive Representation Distillation},
  author={Yang, Puning and Huang, Huaibo and Wang, Zhiyong and Yu, Aijing and He, Ran},
  booktitle={Proceedings of the Asian Conference on Computer Vision},
  pages={39--55},
  year={2022}
}

@article{wang2020minilm,
  title={Minilm: Deep self-attention distillation for task-agnostic compression of pre-trained transformers},
  author={Wang, Wenhui and Wei, Furu and Dong, Li and Bao, Hangbo and Yang, Nan and Zhou, Ming},
  journal={Advances in Neural Information Processing Systems},
  volume={33},
  pages={5776--5788},
  year={2020}
}

@article{khalid2021fakeavceleb,
  title={FakeAVCeleb: A novel audio-video multimodal deepfake dataset},
  author={Khalid, Hasam and Tariq, Shahroz and Kim, Minha and Woo, Simon S},
  journal={arXiv preprint arXiv:2108.05080},
  year={2021}
}

@article{korshunov2018deepfakes,
  title={Deepfakes: a new threat to face recognition? assessment and detection},
  author={Korshunov, Pavel and Marcel, S{\'e}bastien},
  journal={arXiv preprint arXiv:1812.08685},
  year={2018}
}

@inproceedings{sanderson2009multi,
  title={Multi-region probabilistic histograms for robust and scalable identity inference},
  author={Sanderson, Conrad and Lovell, Brian C},
  booktitle={International conference on biometrics},
  pages={199--208},
  year={2009},
  organization={Springer}
}

@article{dolhansky2020deepfake,
  title={The deepfake detection challenge (dfdc) dataset},
  author={Dolhansky, Brian and Bitton, Joanna and Pflaum, Ben and Lu, Jikuo and Howes, Russ and Wang, Menglin and Ferrer, Cristian Canton},
  journal={arXiv preprint arXiv:2006.07397},
  year={2020}
}

@inproceedings{zhang2017s3fd,
  title={S3fd: Single shot scale-invariant face detector},
  author={Zhang, Shifeng and Zhu, Xiangyu and Lei, Zhen and Shi, Hailin and Wang, Xiaobo and Li, Stan Z},
  booktitle={Proceedings of the IEEE international conference on computer vision},
  pages={192--201},
  year={2017}
}

@article{hashmi2025avtenet,
  title={AVTENet: A Human-Cognition-Inspired Audio-Visual Transformer-Based Ensemble Network for Video Deepfake Detection},
  author={Hashmi, Ammarah and Shahzad, Sahibzada Adil and Lin, Chia-Wen and Tsao, Yu and Wang, Hsin-Min},
  journal={IEEE Transactions on Cognitive and Developmental Systems},
  year={2025},
  publisher={IEEE}
}

@inproceedings{rossler2019faceforensics++,
  title={Faceforensics++: Learning to detect manipulated facial images},
  author={Rossler, Andreas and Cozzolino, Davide and Verdoliva, Luisa and Riess, Christian and Thies, Justus and Nie{\ss}ner, Matthias},
  booktitle={Proceedings of the IEEE/CVF international conference on computer vision},
  pages={1--11},
  year={2019}
}

@inproceedings{chugh2020not,
  title={Not made for each other-audio-visual dissonance-based deepfake detection and localization},
  author={Chugh, Komal and Gupta, Parul and Dhall, Abhinav and Subramanian, Ramanathan},
  booktitle={Proceedings of the 28th ACM international conference on multimedia},
  pages={439--447},
  year={2020}
}

@article{ilyas2023avfakenet,
  title={AVFakeNet: A unified end-to-end Dense Swin Transformer deep learning model for audio--visual deepfakes detection},
  author={Ilyas, Hafsa and Javed, Ali and Malik, Khalid Mahmood},
  journal={Applied Soft Computing},
  volume={136},
  pages={110124},
  year={2023},
  publisher={Elsevier}
}

@article{zhang2023joint,
  title={Joint Audio-Visual Attention with Contrastive Learning for More General Deepfake Detection},
  author={Zhang, Yibo and Lin, Weiguo and Xu, Junfeng},
  journal={ACM Transactions on Multimedia Computing, Communications and Applications},
  year={2023},
  publisher={ACM New York, NY}
}

@ARTICLE{10609529,
  author={Wang, Rui and Ye, Dengpan and Tang, Long and Zhang, Yunming and Deng, Jiacheng},
  journal={IEEE Signal Processing Letters}, 
  title={AVT$^{2}$-DWF: Improving Deepfake Detection With Audio-Visual Fusion and Dynamic Weighting Strategies}, 
  year={2024},
  volume={31},
  number={},
  pages={1960-1964},
  doi={10.1109/LSP.2024.3433596}}

@inproceedings{cozzolino2023audio,
  title={Audio-visual person-of-interest deepfake detection},
  author={Cozzolino, Davide and Pianese, Alessandro and Nie{\ss}ner, Matthias and Verdoliva, Luisa},
  booktitle={Proceedings of the IEEE/CVF conference on computer vision and pattern recognition},
  pages={943--952},
  year={2023}
}

@article{usmani2025spatio,
  title={Spatio-temporal knowledge distilled video vision transformer (STKD-VViT) for multimodal deepfake detection},
  author={Usmani, Shaheen and Kumar, Sunil and Sadhya, Debanjan},
  journal={Neurocomputing},
  volume={620},
  pages={129256},
  year={2025},
  publisher={Elsevier}
}

@article{hsu2021hubert,
  title={Hubert: Self-supervised speech representation learning by masked prediction of hidden units},
  author={Hsu, Wei-Ning and Bolte, Benjamin and Tsai, Yao-Hung Hubert and Lakhotia, Kushal and Salakhutdinov, Ruslan and Mohamed, Abdelrahman},
  journal={IEEE/ACM transactions on audio, speech, and language processing},
  volume={29},
  pages={3451--3460},
  year={2021},
  publisher={IEEE}
}

@inproceedings{chollet2017xception,
  title={Xception: Deep learning with depthwise separable convolutions},
  author={Chollet, Fran{\c{c}}ois},
  booktitle={Proceedings of the IEEE conference on computer vision and pattern recognition},
  pages={1251--1258},
  year={2017}
}

@article{chung2018voxceleb2,
  title={Voxceleb2: Deep speaker recognition},
  author={Chung, Joon Son and Nagrani, Arsha and Zisserman, Andrew},
  journal={arXiv preprint arXiv:1806.05622},
  year={2018}
}

@ARTICLE{8449842,

  author={Zhang, Kaihao and Luo, Wenhan and Zhong, Yiran and Ma, Lin and Liu, Wei and Li, Hongdong},

  journal={IEEE Transactions on Image Processing}, 

  title={Adversarial Spatio-Temporal Learning for Video Deblurring}, 

  year={2019},

  volume={28},

  number={1},

  pages={291-301},

  doi={10.1109/TIP.2018.2867733}}

@article{wang2025adaptive,
  title={Adaptive Low Light Enhancement via Joint Global-Local Illumination Adjustment},
  author={Wang, Haodian and Song, Yaqi},
  journal={arXiv preprint arXiv:2504.00400},
  year={2025}
}

@inproceedings{hou2024polyglotfake,
  title={Polyglotfake: A novel multilingual and multimodal deepfake dataset},
  author={Hou, Yang and Fu, Haitao and Chen, Chunkai and Li, Zida and Zhang, Haoyu and Zhao, Jianjun},
  booktitle={International Conference on Pattern Recognition},
  pages={180--193},
  year={2024},
  organization={Springer}
}

@article{zhang2025lightweight,
  title={Lightweight Joint Audio-Visual Deepfake Detection via Single-Stream Multi-Modal Learning Framework},
  author={Zhang, Kuiyuan and Pei, Wenjie and Lan, Rushi and Guo, Yifang and Hua, Zhongyun},
  journal={arXiv preprint arXiv:2506.07358},
  year={2025}
}

@article{zhang2022deep,
  title={Deep image deblurring: A survey},
  author={Zhang, Kaihao and Ren, Wenqi and Luo, Wenhan and Lai, Wei-Sheng and Stenger, Bj{\"o}rn and Yang, Ming-Hsuan and Li, Hongdong},
  journal={International Journal of Computer Vision},
  volume={130},
  number={9},
  pages={2103--2130},
  year={2022},
  publisher={Springer}
}

@inproceedings{campbell2000query,
  title={Query learning with large margin classifiers},
  author={Campbell, Colin and Cristianini, Nello and Smola, Alex and others},
  booktitle={ICML},
  volume={20},
  number={0},
  pages={0},
  year={2000}
}

\end{document}